\documentstyle[12pt]{article}
\begin{document}

\baselineskip=20pt

\title{Dimension minimization of a quantum automaton}

\author{A. M. Martins\\
CFP, Instituto Superior T\'{e}cnico, 1049-001
Lisboa, Portugal }

\date{}

\maketitle

\begin{abstract}

A new model of a Quantum Automaton (QA),  working with qubits is proposed. The quantum states of the automaton can be {\it pure} or  {\it mixed}  and are represented by {\it density operators}. This is the appropriated approach to deal with measurements and dechorence.. 

The linearity of a QA and of the {\it partial trace} super-operator, combined with the properties of invariant subspaces under unitary transformations, are used to minimize the dimension of the automaton and, consequently, the number of its working qubits. 

The results here developed are valid wether the state set of the QA is finite or not. 

There are two main results in this paper: 1) We show that the dimension reduction is possible whenever the unitary transformations, associated to each letter of the input alphabet, obey a set of conditions. 2) We develop an algorithm to find out the equivalent minimal QA and prove that its complexity is polynomial in its dimension and in the size of the input alphabet.

\vskip 2cm \noindent {\bf Draft paper: not for distribution}

\end{abstract}

\newpage

\newpage  

\section{Introduction}

In order to understand computation in a quantum context, it might be useful to introduce as many concepts as possible from the classical computation theory to the quantum case.

One of these basic concepts concerns the functioning of finite automata in classical systems. To extend these concepts from classical to quantum systems, several models have been proposed \cite{ MOORE, WATROUS, NISHIMURA, AMBAINIS02}, most of them dealing with language recognition. This kind of automata is usually referred to as { \it acceptors}, in the sense that they are designed to indicate, by giving a {\it yes} or {\it no} output, whether a given input sequence does or does not possess the proper characteristics.

In this article we propose a new model for a quantum automaton which acts as a { \it transducer} \cite{BOOTH}. This QA is prepared in a given initial quantum state,  then a classical device reads an input string of letters and apply the corresponding transformations to the quantum part. After all the letters in the string are read, the QA is observed (measured). Finally, the output of the quantum measurement is written in an exterior classical  "tape". In this sense, our QA is similar to the {\bf measured-once} Quantum Finite Automata (QFA) model introduced by \cite{ MOORE}, where the measurements are made only after a sequence of letters is read, and it differs from the {\bf measured-many} QFA model introduced by \cite{WATROUS, WATROUS02} that is observed after reading each letter. Since the outputs of this QA are probabilities, it can be seen as some kind of extension of the probabilistic transducers \cite{PAZ,RABIN}. Let us stress that, our QA does not have necessarily a finite state set and it is not used for purposes of language recognition. One important application of transducers is in the measurement, control and identification of systems. 

The study of such problems starts with the minimal-state equivalent automaton which suggests us that an important issue to take into account, when a given QA is defined, is how to minimize it. As far as we know such a question was never addressed in the context of QA. 

The main purpose of this article is to study the minimization procedure of a QA which differs from the procedure used with classical automata.

The physical support of the QA that we are proposing in this paper is a system of $n$ two-state quantum particles which encode information in the form of quantum bits, {\it qubits}. Each qubit is a unit vector of a 2-dimensional Hilbert space. Beside the formal reasons mentioned above, the minimization of a QA is an important question to be addressed because qubits are a very expensive resource. 

The minimization problem can be studied in two complementary ways. The first one, concerns the minimization of the dimension of the underlying Hilbert space, which consists in looking for an equivalent automaton working with the smallest number of qubits. Once the dimension minimization is achieved, another question can be asked, namely the possibility of minimizing the cardinality of the QA state set.

Most authors represent the states of a quantum automaton by {\it pure states}. In general,  a quantum system is not in a pure state. This may be due to the fact  that we have only partial information about the system, or that the system is not isolated from the rest of the Universe, so it is not in a well defined pure state. In such cases we say that the system is in a { \it mixed state} which is represented by a {\it density operator}. The density operator is the most general and powerful way of expressing the state of a quantum system, namely when quantum measurements are performed \cite{KITAEV, AMBAINIS, CAVES2000}, therefore we adopt this formalism to describe the states of a quantum automata.  

Another advantage of describing the states of a QA by density operators, is to deal with quantum measurements performed in subsystems of a larger quantum system. The approach developed in this paper to minimize the dimension of the QA uses the operation of {\it tracing out} that transforms density operators of a Hilbert space in density operators of another Hilbert space of smaller dimension. The transformed density operators are named { \it reduced density operators }.

A QA is a special case of automata, this is, a QA is a linear automaton since its quantum states are vectors of a Hilbert space and the transition and output maps are linear transformations. Based on this linearity, we apply some known results of the theory of invariant subspaces to derive the necessary and sufficient conditions that the unitary evolution operators must obey, in order to minimize its dimension.

This article is organized as follows: in section 2,  the physics background is presented and the physical notation, used in quantum systems, is introduced. In section 3, the definition of our quantum automaton  is presented and its functioning is explained. The conditions for reduction of its dimension are derived in section 4. The minimization algorithm is developed and its complexity is computed in section 5, . Finally, the conclusions are presented in section 6.

\section{Physics background}

\bigskip

A quantum physical system in a { \it  pure state} is described by a unit vector in a Hilbert space, i.e., a vector space with a inner product. In the Dirac notation, the pure state is denoted by $ \mid \psi \rangle $. The quantum automaton, that we are considering, works with n-qubits which can be physically realized by $ n $ two-state particles. The Hilbert space of such a system is $ {\cal H}_2^{n} = { \cal C }^{ \{ 0,1\}^{n} } $, this is a $ 2^n$ dimensional complex vector space. $ {\cal H}_2^{n} $ is a tensor product of $n $ Hilbert spaces, each one associated to a qubit: $ {\cal H}_2^{n} = \bigotimes_{j}^{n} {\cal H}_2(j) $ where ${\cal H}_2(j)$ is the 2-dimensional Hilbert space of qubit $j$. For each $ {\cal H}_2(j)$ we choose a special basis set, called {\it computational basis}, consisting of two orthonormal states $ \mid i_j \rangle , (i_j =0,1) $. A basis set for $ {\cal H}_2^{n} $  is $ \{ \bigotimes_{{j}=1}^{n} \mid i_j \rangle , i_j=0,1 \} $. A general pure state of n-qubits is a vector superposition of the type: $\mid \Psi \rangle = \sum_{i_j,...,i_n = 0}^{1} \psi_{i_{1}....i_{n}}  \bigotimes_{j=1}^{n} \mid i_j \rangle  $, where $ \sum_{i_j,...,i_n = 0}^{1} \mid \psi_{i_{1}....i_{n}} \mid ^{2} = 1$. The transposed-complex conjugate of $ \mid \Psi \rangle $ is denoted by $\langle \Psi \mid $. The inner product between $\mid \Psi \rangle $ and $\mid \Phi \rangle $ is denoted by $ \langle \Phi \mid \Psi \rangle = \sum_{i_1,...,i_n=0}^{1} \psi_{i_1,...,i_n} \phi_{i_1,...,i_n}^{*}$. We define $\mid \Psi \rangle \langle \Phi \mid$ to be the linear operator from ${\cal H}_2^{n}  \longrightarrow {\cal H}_2^{n}$, known as the {\it outer product} of two vectors of ${\cal H}_2^{n}$,  whose matricial representation, in the above mentioned computational basis, is $  \bigotimes_{{j}=1}^{n} \bigotimes_{{k}=1}^{n} \langle i_j \mid  \Psi \rangle \langle \Phi \mid  \mid i_k \rangle  $.

According to the postulates of quantum mechanics, the operations that we can perform in a quantum system are represented by {\bf completely positive maps} \cite{ KITAEV, PRESKILL, KRAUS}. To our QA we are going to apply: a) {\bf unitary transformations}, represented by unitary operators ${ \bf U}$ acting in ${\cal H}_2^{n}$: $ { \bf U} : \mid \Psi \rangle \rightarrow  \mid \Psi  \rangle {'}  = { \bf U } \mid \Psi  \rangle$. The unitary transformations preserve the norm of the vectors in $ {\cal H}_2^{n} $. b) {\bf  measurements of a given physical quantity $ \cal A$}  \cite{COHEN92}. Such a quantity is represented by an hermitian operator $ {\bf A }$ in ${\cal H}_2^{n} $, named { \it  observable quantity}. The measurements can be performed in one or more qubits. The 
possible outcomes of these measurements are given by the eigenvalues $a_k$ (always real) of the operator ${ \bf A}$. The probability of the outcome $a_k$ is  ${ \cal P}_k= Tr \{{ \bf P}_k \mid \Psi  \rangle \langle \Psi \mid \} $, where ${ \bf P}_k= \sum_{i=1}^{g_k}  \mid a_k^{i} \rangle \langle a_k^{i}  \mid $, is the projection operator in the $ g_k$ degenerated eigenspace ${ \cal E}_k$, associated to the eigenvalue $a_k$. The set of vectors, $ \{ \mid a_k^{i} \rangle\}$, constitute an orthonormal basis set in $ {\cal H}_2^{n}$, obeying the eigenvalue equation: $ {\bf A }\mid a_k^{i} \rangle = a_k \mid a_k^{i} \rangle;  0<k<n ;  0< i< g_k  $,  and $ Tr $ stands for the trace of an operator. After a measurement, the quantum state of the system changes according to the outcome $a_k$ becoming, $ \mid  \Psi^{'} \rangle = { \bf P}_k \frac{ \mid \Psi  \rangle}{\sqrt {\langle \Psi \mid P_k  \mid \Psi \rangle}}$.

We have already mentioned that the most general state of a quantum system is not a {\it pure state}  but rather a {\it mixed state}. We say that the system is in a { \it mixed state}, and assign with the system a probability distribution $ \{ p_r,  \mid \psi_r  \rangle, \sum_{r} p_r =1  \} $, meaning that the system is in the pure state $ \mid \psi_r \rangle $, with the probability $ p_r$. A straightforward way of describing such a state is by using density operators $\rho$ defined by \cite{COHEN92}, $ \rho = \sum_{r}  p_r \mid \psi_r \rangle \langle \psi_r \mid$.
The density operators are: {\bf a.} Linear operators of ${\cal H}_2^{n} \rightarrow {\cal H}_2^{n} $;  {\bf b.} Hermitian, $ \rho = \rho^{\dag } $; {\bf c.} Semi definite positive, i.e., for any vector $ \mid \Psi \rangle \in {\cal H}_2^{n} $,  $  \langle  \Psi  \mid  \rho  \mid \Psi \rangle \geq 0 $; {\bf d.} The trace, $ Tr \{ \rho \} = \sum_k \sum_{i}  \langle a_k^{i}\mid \rho  \mid a_k^{i} \rangle = 1$.

The set ${\cal L}_ {{\cal H}_2^{n} }$, of all linear operators of ${\cal H}_2^{n}  \longrightarrow {\cal H}_2^{n}$ is a vector space. Given any two vectors ${\bf A}$ and ${\bf B}$ of ${\cal L}_ {{\cal H}_2^{n} }$ we can define a inner product function by $ ({\bf A},{\bf B}) = Tr \{ {\bf A^{\dagger} B }\} $, called the {\it Hilbert-Schmidt inner product}. With this inner product function, the vector space ${\cal L}_ {{\cal H}_2^{n} }$, becomes a Hilbert space. To any unitary operator, $ { \bf U} : \mid \Psi \rangle \rightarrow  \mid \Psi  \rangle {'}  = { \bf U } \mid \Psi  \rangle$ on ${\cal H}_2^{n}$, we can assign another unitary operator, $ { \bf {\bar U}}: {\bf A} \rightarrow {\bf {\tilde A}} = { \bf {\bar U}}( {\bf A})=  {\bf UAU}^{\dagger}$ on ${\cal L}_ {{\cal H}_2^{n} }$. Operators like $ { \bf {\bar U}}$, that transform operators into other operators are called {\it super-operators}.

An important application of the density operators formalism is as a descriptive tool for subsystems of a composite quantum system. In fact, most physical systems are constituted by two or more parts (subsystems). Let us assume that the observable quantity $ { \bf A}(1 )$ measures only part (1), of a system, for instance the part composed by the $n_1$ first qubits and call part (2) to the remaining $n_2$ qubits, this is $ {\cal H}_2^{n} = {\cal H}_2^{n_1} \otimes {\cal H}_2^{n_2}$.  

The observable quantities $ {\bf A }(1) $  on a subsystem (1) can be extended to a composite system (1) + (2) as follows  \cite{COHEN92}: $ { \bf \tilde {A}} (1) =  {\bf A }(1) \otimes  {\bf 1} (2)$, where ${ \bf {\tilde A} }(1) $ denotes the corresponding observable for the same physical measurement, performed on the composite system and ${ \bf  1}(2)$ is the identity operator in ${\cal H}_2^{n_2}$. The action of this operator on the vector, $ \mid \Psi (1) \rangle \otimes \mid  \Psi (2) \rangle$ is defined by,
\begin{equation}\label{extension}
{ \bf \tilde {A}} (1) \left[ \mid \Psi (1) \rangle \otimes \mid \Psi (2) \rangle \right]=  \left[  {\bf A } (1 ) \mid \Psi (1) \rangle \right] \otimes  \left[{ \bf 1} (2) \mid \Psi (2) \rangle \right]
\end{equation}

The spectrum of  ${ \bf \tilde {A}} (1)$ in $ {\cal H}_2^{n}  $ is the same as the spectrum of $ {\bf {A}} (1)$ in ${\cal H} _2^{n_1}$, but with all the eigenvalues degenerated in $ {\cal H}_2^{n}$, even if none of them is degenerated in ${\cal H}_2^{n_1}$. The probability $ {\cal P}_k $ of obtaining the outcome $a_k$, when the observable ${ \bf  \tilde A } (1)$ is measured over part (1) of the system is given by ${\cal P}_k= Tr \{{\bf \tilde {P}_k}(1) \rho  \}$, where $ {\bf \tilde {P}_k}  (1)= {\bf P}_k (1) \otimes { \bf 1 } (2) $ and $ {\bf P}_k (1)$ is the projector operator on the eigenspace $ {\cal E}_k $. 

A natural question we can ask is, would it be possible to define a density operator for subsystem (1), that gives the correct probabilities $ { \cal P}_k $ for the outcomes of the observable $ { \bf A}(1)$? 

It can be shown \cite{PRESKILL, KRAUS} that there is a unique transformation of ${\cal L}_ {{\cal H}_2^{n} }$ into  ${\cal L}_ {{\cal H}_2^{n_1 } }$ which gives rise to the correct description of observable quantities for subsystem (1). This transformation, called { \it trace out } or {\it partial trace} and denoted $Tr_2 \{ . \} $, is a {\it completely positive map} of ${\cal  L}_{{{\cal H}_2^{n}},{{\cal H}_2^{n_1}}} $ which is the space of all linear operators: $ {\cal L}_{{\cal H}_2^{n}}  \longrightarrow   {\cal L}_{{\cal H}_2^{n_1}} $. It is defined for any pair of finite dimensional Hilbert spaces $ {\cal  L}_{{\cal H}_2^{n} }$ and $ {\cal L}_{{\cal H}_2^{n_1}}$, with  $n_1<  n$. The  {\it image} of the partial trace of a density operator $\rho \in {\cal L}_{{\cal H}_2^{n }}$ is denoted by $ \rho^1 = Tr_2 \{\rho \}$ and is named {\it reduced density operator} for subsystem (1), whose matrix elements are $\rho^1 (i,j) =  {\sum_{k=n_1 +1}^{n}} \langle k | \rho | k  \rangle = \sum_{k=n_1 +1}^{n} \rho(ik,jk)$. It means that we are averaging over $ {\cal H}_2^{n_2} $. Any quantum operator that does not operate on ${\cal H}_2^{n_2} $ commutes with the partial trace. The reduced density operators have the same properties as the density operators. Let (A) to be a subsystem of $n_A$ qubits, containing the first $n_1$ qubits and let (B) to be the remaining $n_B$ qubits, then $n_A=n-n_B$. The partial trace has the following property:
\begin{equation}\label{reduced}
{ \cal P}_k = Tr\{{\bf {\tilde P}_k} (1)  \rho \}=Tr_A \{ {\bf { \tilde P}_k }(1) \rho^{A} \}= Tr_1 \{ {\bf P_k} (1) \rho^1 \}.
\end{equation}
Where $ \rho^{A} = Tr_B \{\rho \}$ is the reduced density operator of subsystem (A) and $Tr_B \{.\}$ is the partial trace over subsystem (B). This equality shows that it is enough to know the reduced density operator $\rho^1$ (or $\rho^{A}$) in order to compute the probabilities of all the outcomes of a measurement over subsystem (1).

\section{ A quantum automaton }

\bigskip

We say that a quantum system of $n$ qubits works as a quantum automaton $ {\cal M}= \langle {\cal H}_2^{n}, \rho_0, \Sigma,  {\cal U}, Q, {\bf {\tilde A}}(1), \Omega \rangle $ when the following conditions are fulfilled:

{\it 1.  ${\cal H}_2^n$ is the underlying Hilbert space of dimension $2^n$.

2. $\rho_0 \in {\cal L}_{{\cal H}_2^{n}}$ is the density operator of the initial quantum state of the $n$ qubits.

3. $\Sigma=\{\sigma \}$ is a finite set of input symbols (the input alphabet).\

4. $Q$ is the set of reachable states given by:
$Q= \{ \rho_w : w \in \Sigma^{*} \}$, with $\rho_w= {\bf {\bar U}}_w (\rho_0 )= {\bf U}_w \rho_0 {\bf U}_w^{\dag } $ and where the family $\{ { \bf  U}_w \}_{w \in \Sigma^*}$ of unitary operators of ${\cal H}_2^n$ is built as follows:  i) ${\cal U}= \{{\bf U}_{\sigma} \}_{\sigma \in \Sigma}$, is a finite set of unitary operators indexed upon $ \sigma \in \Sigma  $; ii) ${ \bf  U}_{\epsilon}={\bf 1}(n) $;  and iii) ${\bf U}_{w \sigma}= {\bf U}_{\sigma} {\bf U}_w$.

5. $ \delta: \Sigma \times  Q \longrightarrow Q$ is the transition map defined by
\begin{equation}
 \delta (\sigma ,\rho_w) =  \rho_{w \sigma} ={ \bf {\bar U}}_{\sigma} ( \rho_w )
\end{equation}

6. ${\bf {\tilde A}}(1)$ is the observable quantity to be measured. The possible outcomes of a measurement of ${\bf {\tilde A}}(1)$ are its eigenvalues $a_k$: $ Spec {\bf {\tilde A}}(1) =\{ a_k : {\bf {\tilde A}}(1)( | a_k^{i_k} \rangle \otimes  {\bf 1}(2) )=a_k ( | a_k^{i_k} \rangle \otimes  {\bf 1}(2) ) \} $. The diagonal representation of the observable is $ {\bf {\tilde A}}(1) =\sum_{k} a_k {\bf {\tilde P}_k}(1)$, where ${\bf {\tilde P}_k}(1)= \sum_{i_k=1}^{g_k}  \mid a_k^{i_k} \rangle \langle a_k^{i_k}  \mid\otimes  {\bf {\bar 1}} (2)  $ is the projector into the eigenspace ${\cal E}_k$ of ${\bf {\tilde A}}(1)$.

7. $\Omega $ is a set of output symbols (the output alphabet) whose elements are defined by the following output map:

8. $\omega: Q \rightarrow \Omega $ 
\begin{equation}\label{propability}
\omega ( w, a_k) = Tr \{ { \bf {\tilde P}_k}(1) \rho_w  \} 
\end{equation}
For each $w \in \Sigma^{*}$, the set  $\Omega_w= \{\omega ( w, a_k):  a_k \in Spec {\bf {\tilde A}}(1) \}$  is the unique probability measure for the state $\rho_w$ and the output alphabet, $\Omega = \{ \Omega_w, w \in \Sigma^{*}; \,\ \sum_k  \omega ( w, a_k) =1 \}$, is a set of probability measures. } 

The physical performance of this QA is as follows: a classical exterior device can read the symbols $\sigma \in \Sigma$ stored in an external classical "tape".  Then it applies the corresponding gate, represented by the unitary operator ${ \bf U}_{\sigma}$, to the quantum state $\rho_w \in  {\cal L}_{{\cal H}_{2}^n}$, of the $n$ two-state particles where the amount of information is stored. The dimension of the QA ${\cal M}$, denoted by $dim[{\cal M}]$, is the dimension of its quantum states $\rho_w $: $dim[ {\cal M}]= 2^n \times 2^n =N$. After each input string of lenght $w$, the observable ${\bf {\tilde A}}(1)$ is measured and the corresponding output is given by the probability $\omega ( w, a_k) = Tr \{ { \bf {\tilde P}_k}(1) \rho_w  \} $, which is stored in an exterior classical device. 

The concepts of {\it equivalent states} and of {\it equivalent automata} can also be defined for a QA as follows.

Two states  $ \rho_{w_i}$ and $ \rho_{w_j}$ of the state set $Q$ of a QA $ {\cal M}$, are said to be equivalent, and we write $\rho_{w_i} \simeq \rho_{w_j}$ iff  $\omega ({\bf {\bar U}_w}( \rho_{w_i}),a_k )= \omega ({\bf {\bar U}_w} (\rho_{w_j} ), a_k)$ for every $w \in \Sigma^{*}$, More precisely, two states $\rho_{w_{i}}, \rho_{w_{j}} \in Q$ are equivalent, iff for every $ a_k \in Spec {\bf {\tilde A}}(1) $ and for every $w \in \Sigma^{*}$: 
\begin{equation}\label{equstate1}
Tr\{ \rho_{w_{i}} { \bf {\tilde P}_k}(1)  \} = Tr\{ \rho_{w_{j}} { \bf {\tilde P}_k}(1)  \}
\end{equation}
\begin{equation}\label{equstate2}
Tr\{ {\bf {\bar U}_w}( \rho_{w_{i}}) { \bf {\tilde P}_k}(1)  \} = Tr\{ {\bf {\bar U}_w} (\rho_{w_{j}}) { \bf {\tilde P}_k}(1)  \}. 
\end{equation}
We define a { \it quantum behavior} of a QA $ {\cal M}$ over the signature $ (\Sigma, \Omega) $, as a map $ \beta_M(w): \Sigma^{*} \rightarrow \Omega ( \rho_w) $ for every $w \in \Sigma^{*}$. Furthermore, two QA ${\cal M}$ and ${\cal M}_A$,  over the same signature are said to be {\it equivalent} and we write ${\cal M} \simeq {\cal M}_A$, when they have the same observable quantity ${\bf {\tilde A}}(1)$ and the same behavior. The equivalence between quantum automata is a bit more restrictive than between classical ones. The reason is because the output alphabet of a QA is determined by the probabilities of the outcomes of the measurements of an observable quantity ${\bf {\tilde A}}(1)$, therefore all equivalent QA must have in common this observable. In the present case, where the QA works with qubits,  we cannot replace it by another QA with the same signature, if it doesn't work with qubits. Moreover, there is no QA equivalent to ${\cal M}$ with less than $n_1$ qubits, since the observable quantity must always be ${\bf {\tilde A}}(1)$.

  The fact that the dimension of the underlying Hilbert space of a QA is finite doesn't imply that the cardinality, or size, of its state set $Q$ is finite. When the cardinal of $Q$ is finite we call the QA a {\it Quantum Finite Automaton}. In such a case the output alphabet is also finite.
 
The unitary operators ${\bf U}_{\sigma}$ should have some properties in order that a QA might be also a QFA.
 
{\bf Proposition 1:} {\it The number of states $\rho_l ={\bf U}^l \rho_0 {\bf U}^{l \dagger}$ ( $l \in N$), generated by a unitary operator $ {\bf U} $ of ${\cal H}_2^n$, is finite iff there is a $p \in N$ and $p \leq 2^n \times 2^n$, such that $ {\bf U}^p = {\bf 1} $. In this case the unitary operator ${\bf  U} $ has $p$ different imaginary eigenvalues $ u_j = e^{i 2 \pi j /p} $, with $1 \leq j \leq p$, and the set generated by ${\bf U}$ is a finite group.}

{\bf Proof }: The finite number of states of the QA is the result of the well known properties of any group of cyclic unitary operators.
{\bf Q.E.D.}
 
{\bf Corollary 1:} {\it Let $ {\cal M}$ be a QA. If there is at least a $\sigma \in \Sigma $ such that the set generated by the unitary operator $ {\bf U}_{\sigma}$ is not a finite group then the QA ${\cal M}$ is not finite.}

{\bf Corollary 2:} {\it If for each $\sigma \in \Sigma$ the set generated by $ {\bf U}_{\sigma}$ is finite  and if: ${\bf U}_{\sigma '}{\bf U}_{\sigma }={\bf U}_{\sigma }{\bf U}_{\sigma ' }$ for every $\sigma , \sigma ' \in \Sigma $, then $\{ { \bf U}_w \}_{w \in \Sigma^*}$ is a finite commutative group, and the QA $ {\cal M}$ is finite.}

\section{Dimension reduction }

Let us consider a QA, $ {\cal M}= \langle {\cal H}_2^{n}, \rho_0, \Sigma, {\cal U}, Q, { \tilde {\bf A}}(1) , \Omega \rangle$, composed by two subsystems (1) and (2) working with $n_1 $  and $n_2$ qubits, respectively. The total number of qubits  of the QA ${\cal M}$ is $ n=n_1 +n_2 $. The observable quantity of the QA is ${ \bf \tilde A } (1)= {\bf A} (1) \otimes { \bf 1} (2) $ . Measuring this physical quantity corresponds to  "inquiring" only about part (1) of the total system. A natural question we can ask  is: In which conditions would it be possible to replace the QA $ {\cal M}$ by another equivalent QA $  {\cal M}_A $, using only $ n_A $ ($ n_1 \leq n_A < n $ ), qubits? Such a QA would obviously have a smaller dimension than the original one. Moreover, if  $n_A={ \bar n} $ is the least number of qubits such that $ {\cal M}_A  \simeq  { \bar  {\cal M}} $ is equivalent to $ {\cal M}$ then, ${ \bar  {\cal M}} $ is the automaton with the minimal dimension or saying in other words, ${\bar { \cal  M}}$ is the QA  working with the minimal number of qubits.

What is usually understood by the minimization process of a classical automaton is the minimization of the cardinality of its sate set. However, this question can only be addressed if the QA is also a QFA. Instead, we are going to show that given a QA $ {\cal M}$, with a finite or infinite number of states, it is always possible to determine whether its dimension is minimal.

A QA is a special case of automata, this is, a QA is a linear automaton since its states are vectors in a Hilbert space and the transition and output maps are linear transformations. One of the distinct advantages of linear automata  is that their behavior can be studied by analytical rather than the ennumerative techniques used in general automata.

The procedure to minimize the cardinality of a QFA follows the usual approach developed for classical automata, which consists in looking for equivalent states and the corresponding equivalence classes, however there are some specificities for quantum systems that we will present in a future work.

\subsection{Disentangled subsystems}

It was mentioned in section 2, that the unique operation on a system that gives rise to the correct description of an observable quantity referred to one of its subsystems, is the partial trace over the remaining subsystems. This property will be the keystone in the process of dimension reduction of a QA.

{ \bf Theorem 1}: {\it Let $ {\cal M}= \langle {\cal H}_2^{n}, \rho_0, \Sigma,  {\cal U}, Q, {\bf {\tilde A}}(1), \Omega \rangle $ be a QA whose initial state $ \rho_0 $ can be expressed in the form $ \rho_0 = \rho_0^{A} \otimes \rho_0^{B}$, where $ \rho_0^{A}$ is the reduced density operator of part (A) of the system, working with $ n_A $ qubits and containing subsystem (1), and $\rho_0^{B}$ is the reduced density operator of the remaining part (B) of the system, working with $n_B = n -n_A$ qubits.  If all unitary transformations $ { \bf {\bar U}_{ \sigma}} \in { \bf {\cal U}} $, had the form $ { \bf {\bar U}_{ \sigma } }= { \bf {\bar U }_{ \sigma }^{A}} \otimes { \bf {\bar U} _{ \sigma}^{B}}$ (where ${ \bf {\bar U }_{ \sigma }^{A}}$ and ${ \bf {\bar U }_{ \sigma }^{B}} $ are unitary transformations in ${\cal L}_ {{\cal H}_2^{n_A}}$ and ${\cal L}_ {H_2^{n_B}}$, respectively), then the QA $  {\cal M}_A =  \langle {\cal H}_2^{n_A}, \rho_0^{A}, \Sigma, {\cal U}_A, Q_A, {\tilde {\bf A}}(1), \Omega \rangle $ is equivalent to ${\cal M }$, and \\
1 - $ {\cal H}_2^{n_A}$ is the underlying Hilbert space with the minimal dimension $ 2^{ n_A}$.\\
2 - $ \rho_0^{A}= Tr_B\{ \rho_0 \}$ is the initial state.\\
3 - The input alphabet, $\Sigma = \{ \sigma \}$, is the same as in QA ${\cal M}$.\\
4- $ Q_A = \{ \rho_w^{A} : w \in \Sigma^{*} \} $ with $ \rho_w^{A} = Tr_B \{ \rho_w \} =  {\bf { \bar U}_w^{A}} ( \rho_0^{A} )= {\bf  U}_w^{A} \rho_0^{A} {\bf  U}_w^{A \dag}$, and $\{ { \bf U}_w^{A} \}_{w \in \Sigma^*}$ is a family of unitary operators of ${\cal H}_2^{n_A}$ built as follows: i) ${ \bf U}_{\epsilon}^{A}={\bf 1}(A) $;  ii) ${\cal U}_A= \{ {\bf U}_{\sigma}^{A} \}_{\sigma \in \Sigma}$, is a finite set of unitary operators indexed upon $ \sigma \in \Sigma  $ and iii) ${\bf U}_{ \sigma w}^{A}= {\bf  U}_{\sigma}^{A} {\bf U}_w^{A}$.\\
5 - $ \delta_A:  \Sigma \times Q_A \rightarrow Q_A $ such that $ \rho_{ \sigma w}^{A} = \delta_A (\sigma, \rho_w^{A} ) ={ \bf {\bar U }_{\sigma }^{A}}( \rho_w^{A} )$ \\ 
6 - The output function $ \omega_A : Q \rightarrow \Omega $ is defined by $ \omega_A (\rho_w^{A},a_k) = Tr_A \{ {\bf { \tilde P}}_k (1) \rho_w^{A} \} = \omega(\rho_w,a_k) $.\\
7 - $\omega_A: Q_A \rightarrow \Omega$ such that $ \omega_A(\rho_w^{A},a_k) = Tr_A \{ {\bf { \tilde P}_k }(1) \rho_w^{A} \}  $ and $ \Omega = \{ \Omega_w; w \in \Sigma^{*} ; \,\  \sum_k  \omega_A ( w, a_k) =1\}$, with $\,\ \Omega_w= \{\omega_A (\rho_w, a_k):  a_k \in Spec {\bf A}\}$, is the same as in QA  ${\cal M}$. }

{\bf Proof}: 
Using the properties of the partial trace operator: \\
$\rho_{\sigma }^{A}  =  Tr_B\{ \rho_{\sigma} \}=Tr_B\{  {\bf {\bar U }_{ \sigma }}(\rho_0) \}= Tr_B \{{ \bf {\bar U }_{ \sigma }^{A}}( \rho_0^{A}){ \bf {\bar U} _{ \sigma}^{B} } (\rho_0^{B})\} ={ \bf {\bar U }_{ \sigma }^{A}}( \rho_0^{A})$. \\
Using eq.(\ref{reduced}), we obtain: $ \omega_A ({ \bf \rho }_w^{A},a_k) = Tr_A \{ {\bf { \tilde P}}_k (1) \rho_w^{A} \} = Tr \{{ \bf  { \tilde P}}_k (1) \rho_w  \} = \omega (\rho_w,a_k) $.\\
{\bf Q.E.D.}\\
{\bf Theorem 1} says that if part (A) is initially disentangled from part (B) of the system and, if the unitary transformations preserve this disentanglement, then each subsystem (A)  and (B) follows its own dynamics behaving as two separated systems. In this situation we can forget subsystem (B) and consider the QA restricted to subsystem (A), because ${\cal M} \simeq {\cal M_A}$.  The equivalent QA ${\cal M_A}$ is named a {\it sober} QA. {\bf Theorem 1} is a sufficient condition for dimension reduction.

In the next sections we are going to use some known properties of invariant subspaces under unitary transformations,  in order to derive the classes of equivalent quantum states. 

\subsection{Linear transformations and invariant subspaces}

Let ${\bf {\bar T}}: {\cal L}_{{\cal H}_2^{n}} \rightarrow {\cal L}_{{\cal H}_2^{n_A}} $ to be the linear transformation defined by $ {\bf {\bar T}} =Tr_B \{. \} $. Let  $  {\cal K}=ker ({\bf {\bar T}})  $ and $ {\cal I}=Im ({\bf {\bar T}}) $, to be the kernel and the image spaces of the transformation ${\bf {\bar T}}$. In Appendix A it is shown that, $ Im({\bf {\bar T}}) \equiv {\cal L}_{{\cal H}_2^{n_A}}$, therefore ${\bf {\bar T}}$ is a linear transformation onto ${\cal L}_{{\cal H}_2^{n_A}}$. Let ${\cal Q} \subset {\cal L}_{{\cal H}_2^{n}}$ be the subspace orthogonal to ${\cal K}$, i.e. ${\cal L}_{{\cal H}_2^{n}}={\cal Q}   \bigoplus  {\cal K}$, then $ dim [{\cal L}_{{\cal H}_2^{n}} ] = dim[ {\cal Q}] +  dim[ {\cal K}]$, where $dim[.]$ means dimension of a vector space. It is also true that $dim[ {\cal L}_{{\cal H}_2^{n}} ] = dim [{\cal I}] + dim[ {\cal K}] $. and we conclude that $ dim[ {\cal Q}] = dim [{\cal I}]$.  Since the spaces ${\cal Q}$ and ${\cal I}$ have the same dimension they are isomorphic ($1^{st}$ theorem of isomorphism). 

Let us denote $dim[ {\cal L}_{{\cal H}_2^{n}} ] =2^n \times 2^n = d$,  $dim[ {\cal Q}]= 2^{n_A} \times 2^{n_A} = q$ and $dim[ {\cal K}]=2^n \times 2^n - 2^{n_A} \times 2^{n_A} =k  $ and let  $ { \cal B}^{\cal Q }= \{ {\bf B_j^Q }; j=1,...,q \}$, and $ { \cal B}^{\cal K} = \{ {\bf B_j^K} ; j=q+1,...,d \}$, be two orthonormal basis sets for subspaces  ${\cal Q}  $ and ${\cal K}  $, respectively. Their basis vectors obey the following orthonormal relations,
 \begin{equation}\label{orthorelations}
({\bf B_i^{Q}} , {\bf B_j^{Q}})= ({\bf B_i^{K}} ,{\bf  B_j^{K}}) = \delta_{ij} \,\,\ ; \,\,\,\
({\bf B_i^{Q}} , {\bf B_j^{K}})=({\bf B_i^{K} }, {\bf B_j^{Q}})=0
\end{equation}
where $(.,.) $ is the Hilbert-Schmidt inner product defined in ${\cal L}_{{\cal H}_2^{n}}$. 

The properties of any unitary transformation ${\bf {\bar U}}$ are completely determined by its effect on the basis-${\cal B^{Q}} \cup {\cal B^{K}}$. Thus, if 
\begin{equation}\label{devel}
{\bf Z} = \sum_{j=1}^{q} z_j {\bf B_j^{Q}} + \sum_{j=1+q}^{d} z_j {\bf B_j^{K}}.
\end{equation}
is any vector in $ {\cal L}_{{\cal H}_2^{ n}}$, the image of ${\bf Z}$ under ${\bf {\bar U}}$ is 
\begin{equation}\label{uz}
{\bf {\bar U}}({\bf Z} )= \sum_{j=1}^{q} z_j {\bf {\bar U}}({\bf B_j^{Q}}) + \sum_{j=1+q}^{d} z_j {\bf {\bar U}}({\bf B_j^{K}})
\end{equation}
therefore, if the images of the basis vectors are known, the image of any vector can be determined. The image of the basis-${\cal B^{Q}} \cup {\cal B^{K}}$, under ${\bf {\bar U}}$ is, 
\begin{equation}\label{ub1}
{\bf {\bar U}} ({\bf B_j^{Q}}) =  \sum_{i=1}^{q}   {\bar u}_{ij}^{QQ} {\bf B_i^{Q} } + \sum_{i=1 + q}^{d}  {\bar u}_{ij}^{QK} {\bf B_i^{K} } \,\,\ ;j=1,...,q
\end{equation} 
\begin{equation}\label{ub2}
{\bf {\bar U}} ({\bf B_j^{K}}) =   \sum_{i=1}^{q}    {\bar u}_{ij}^{KQ} {\bf B_i^{Q} } + \sum_{i=1 + q}^{d}  {\bar u}_{ij}^{KK} {\bf B_i^{K} }\,\,\ ;j=q+1,...,d
\end{equation} 
where,
\begin{equation}\label{block1}
{\bar u}_{ij}^{QQ}= ({\bf B_i^{Q}}, {\bf {\bar U}} ({\bf B_j^{Q}}))= Tr\{ {\bf B_i^{Q}}^{\dag }{\bf {\bar U}} ({\bf B_j^{Q}}) \} \,\,\  ; i=1,...,q; \,\,\  j=1,...,q
\end{equation}
\begin{equation}\label{block2}
{\bar u}_{ij}^{KK}= ({\bf B_i^{K}} ,{\bf {\bar U}} ({\bf B_j^{K}}))=Tr\{ {\bf B_i^{K}}^{\dag } {\bf {\bar U}} ({\bf B_j^{K}}) \} \,\ ; i=q+1,...,d\,\  ; j=q+1,...,d
\end{equation}
\begin{equation}\label{block3}
{\bar u}_{ij}^{QK}= ( {\bf B_i^{K}},{\bf {\bar U}} ({\bf B_j^{Q}}) )=Tr\{{\bf B_i^{K}}^{\dag } {\bf {\bar U}} ({\bf B_j^{Q}}) \} \,\,\ ;  i=q+1,...,d \,\ ; j=1,...,q
\end{equation}
\begin{equation}\label{block4}
{\bar u}_{ij}^{KQ}= ( {\bf B_i^{Q}}, {\bf {\bar U}} ({\bf B_j^{K}}))=Tr\{{\bf B_i^{Q}}^{\dag }{\bf {\bar U}} ({\bf B_j^{K}})  \} \,\,\ ;  i=1,...,q \,\ ; j=q+1,...,d
\end{equation}
are the matrix elements of the super-operator ${\bf {\bar U}}$ in the basis-${ \cal B}^{\cal Q} \cup { \cal B}^{\cal K}$.

Replacing ${\bf {\bar U}}({\bf B_j^{Q}})$ and  $ {\bf {\bar U}}({\bf B_j^{K}})$ given by (\ref{ub1}) and (\ref{ub2}) into eq.(\ref{uz}), we obtain
\begin{equation}\label{UZ}
{\bf {\bar U}}({\bf Z} )= \sum_{j=1}^{q} \{\sum_{j=1}^{q}{\bar u}_{ij}^{QQ}z_j +  \sum_{j=1+q}^{d}{\bar u}_{ij}^{KQ}z_j  \}  {\bf B_j^{Q}} + \sum_{j=1+q}^{d} \{\sum_{j=1}^{q}{\bar u}_{ij}^{QK}z_j +  \sum_{j=1+q}^{d}{\bar u}_{ij}^{KK}z_j  \}  {\bf B_j^{K}} 
\end{equation}
A possible relation between the subspaces ${\cal Q}$ and ${\cal K}$ of ${\cal L}_{{\cal H}_2^n}$ and the linear transformation ${\bf {\bar U}}$, is invariance. We say that ${\cal Q}$ (or ${\cal K}$) is invariant under ${\bf {\bar U}}$ if, for every $ {\bf X} \in {\cal Q}$ (or, for every ${\bf Y} \in {\cal K}$) implies ${\bf {\bar U}}({\bf X}) \in {\cal R} $ (or ${\bf {\bar U}} ({\bf Y}) \in {\cal K} $ ). We also say that ${\cal Q}$ (or $ {\cal K}$) is invariant under a set of linear transformations ${\bf {\bar U}}_{\sigma}$ ($\sigma \in \Sigma$), if it is invariant under each member of the set. When two spaces, say ${\cal Q}$ and ${\cal K}$, such that  ${\cal Q} \bigoplus {\cal K} ={\cal L}_{{\cal H}_2^n}$, are both invariant under ${\bf {\bar U}}$, then we say that ${\bf {\bar U}}$ is reduced (decomposed) by the pair (${\cal Q}, {\cal K}$). The following proposition \cite{HALMOS} defines the structure of the matrix representing a unitary transformation ${\bf {\bar U}}$ under which ${\cal Q}$ (or ${\cal K}$) is invariant.

{\bf Proposition 2: }  {\it Let $ {\cal L}_{{\cal H}_2^{n}} = {\cal Q} \bigoplus {\cal K}$ to be the direct sum of the subspaces ${\cal Q}$ and ${\cal K}$. If ${\cal K}$  (or ${\cal Q}$) is invariant under a unitary operator ${\bf {\bar U}}$, so is the complementary subspace ${\cal Q}$ (or ${\cal K}$). Then, the unitary transformation ${\bf {\bar U}}$ is reduced by the pair $({\cal Q},{\cal K})$ and the matrix representation of ${\bf {\bar U}}$ in the basis-${ \cal B}^{\cal Q } \cup { \cal B}^{\cal K}$ is}
\begin{equation}\label{matrix1}
{\bf  {\bar U} }= 
 \left(
\begin{array}{cc}
{\bf {\bar U}^{QQ} } &   {\bf 0 }   \\
{\bf 0 }          & {\bf {\bar U}^{KK} } 
\end{array}
\right) \
\end{equation}
{\it where the block diagonal matrices 
\begin{equation}\label{block11}
{\bf {\bar U}^{QQ}}= [ {\bar u}_{ij}^{QQ} ]_{i=1,...,q \,\ ;j=1,...,q}
\end{equation}
\begin{equation}\label{block22}
{\bf {\bar U}^{KK}}= [ {\bar u}_{ij}^{KK} ]_{i=q+1,...,d ;j=q+1,...,d}
\end{equation}
are unitary matrices, and the off diagonal blocks are} 
\begin{equation}\label{block12}
{\bf {\bar U}^{QK}}= [ {\bar u}_{ij}^{QK} ]_{i=q+1,...,d \,\ ;j=1,...,q}=0
\end{equation}
\begin{equation}\label{block21}
{\bf {\bar U}^{KQ}}= [ {\bar u}_{ij}^{KQ} ]_{i=1,...,q ;j=q+1,...,d}=0
\end{equation}
Whenever ${\bf  {\bar U} }$ is reduced by the pair $({\cal Q},{\cal K})$ then, ${\bf  {\bar U} }= {\bf  {\bar U}^{QQ} } \bigoplus {\bf  {\bar U}^{KK}}$  is the direct sum of the unitary transformations ${\bf {\bar U}^{QQ}}$ and ${\bf {\bar U}^{KK}}$ defined on the subspaces ${\cal Q}$ and ${\cal K}$, res-pectively. The transformation ${\bf {\bar U}^{QQ}}$ describes what ${\bf  {\bar U} }$ does on ${\cal Q}$ and the transformation ${\bf {\bar U}^{KK}}$  describes what ${\bf  {\bar U} }$ does on ${\cal K}$.

{\bf Proposition 3:} {\it Let ${\cal K} \in  {\cal L}_{{\cal H}_2^{n}} $ to be invariant under the unitary transformations ${\bf {\bar U}}_a $ and ${\bf {\bar U}}_b$. Then ${\cal K}$ is also invariant under the transformation ${\bf {\bar U}}_{a}{\bf {\bar U}}_{b}$.}

{\bf Proof:} The product of diagonal matrices is also a diagonal matrix. {\bf QED}.

What will the action of a unitary super-operator ${\bf {\bar U}}$ on a general vector  ${\bf Z} \in  {\cal L}_{{\cal H}_2^{n}} $ be?

Applying the linear super-operator ${\bf {\bar T}}= Tr_B\{.\}$ to both sides of this eq.(\ref{devel}), and denoting by $ {\bf B_j^{I}}={\bf {\bar T}}( {\bf B_j^{Q}}) ; j=1,...q$, we obtain
\begin{equation}\label{Tdensity}
{\bf Z}^{A}= {{\sum}_{j=1}^{q} }z_j {\bf B_j^{I} } = \sqrt{2^{n-n_A}}{{\sum}_{j=1}^{q} }z_j {\bf B_j^{' I} } 
\end{equation} 
The set $ {\cal B}^{I}= \{ {\bf B_j^{' I}}; j=1,...,q \}$ is an orthonormal basis set of ${\cal L}_{{\cal H}_2^{n_A}} $ (Appendix A), and ${\bf Z }^{A} =Tr_B\{{\bf Z} \} \in {\cal L}_{{\cal H}_2^{n_A}} $ is the image of ${\bf Z}$.

The image of $\bf Z$ under ${\bf {\bar U}}$ is ${\bf {\tilde Z}}= {\bf {\bar U}}({\bf Z})$ given by eq.(\ref{uz}). Applying ${\bf {\bar T}}$ to both sides of it, we end up with,
\begin{equation}\label{ua}
{\bf {\tilde Z}}^{A}=   \sqrt{2^{n-n_A}} \sum_{i=1}^{q} \left( {{\sum}_{j=1}^{q} } {\bar u}_{ij}^{QQ} z_j +{{\sum}_{i=q+1}^{d} } {\bar u}_{ij}^{KQ}  z_j \right){\bf B_i^{' I} } 
\end{equation}
where ${\bf {\tilde Z}}^{A} ={\bf {\bar T}} ({\bf {\tilde Z}})$.

We would like to know if there is a unitary super-operator ${\bf {\bar U}}^{A}: {\cal L}_{{\cal H}_2^{n_A}} \rightarrow {\cal L}_{{\cal H}_2^{n_A}} $, such that for every state ${\bf Z} \in {\cal L}_{{\cal H}_2^n} $:
\begin{equation}\label{condition}
{\bf {\bar U}}^{A} {\bf {\bar T} } \left ({\bf Z} \right) =  {\bf {\bar T}}{\bf {\bar U}}\left ({\bf  Z} \right )
\end{equation}
The action of any super-operator ${\bf {\bar U}}^{A}$ on any ${\bf Z}^{A} \in {\cal L}_{{\cal H}_2^{n_A}} $ is, 
\begin{equation}\label{upa}
{\bf {\bar U}}^{A}( {\bf Z}^{A} )= \sqrt{2^{n-n_A}}{{\sum}_{i=1}^{q} }  \left[ {\sum}_{j=1}^{q}  {\bar u}_{ij}^{A} z_j \right]  {\bf B_i^{' I} } 
\end{equation}
where $ {\bar u}_{ij}^{A}= \left ( {\bf B_i^{' I}}, {\bf {\bar U}}^{A}({\bf B_j^{' I}}) \right )$ are the elements of ${\bf {\bar U}}^{A} $ relative to the ${\cal B}^{I}$-basis. Replacing (\ref{upa}) and (\ref{ua}) in eq.(\ref{condition}) and recalling that ${\cal B}^{I} $ is a set of linearly independent vectors, we obtain
\begin{equation}\label{C1}
{{\sum}_{j=1}^{q} } \left( {\bar u}_{ij}^{A} -  {\bar u}_{ij}^{QQ} \right) z_j - {{\sum}_{j=q+1}^{d} }  {\bar u}_{ij}^{KQ} z_j =0; \,\,\,\,\  \forall_{i=1,...,q} 
\end{equation}
This is a system of $q$ linear equations, where the unknowns are the $q^2$ elements $ {\bar u}_{ji}^{A}$ of matrix ${\bf {\bar U}}^{A} $. There are $(q^2-q)$ linearly independent non-trivial solutions. We can choose freely $(q^2-q)$ numbers ${\bar u}_{ji}^{A}$ and the remaining $q$ values being uniquely determined. Each of these solutions will depend on the components $z_j $ of the vector ${\bf Z}$, except when the coefficients of every $z_j$, in eq.(\ref{C1}), are simultaneously null, i.e., 
\begin{equation}\label{C3}
{\bar u}_{ij}^{A}= {\bar u}_{ij}^{QQ}  \,\,\ i=1,...,q; \,\,\ j=1,...,q  
\end{equation}
\begin{equation}\label{C4}
{\bar u}_{ij}^{KQ} =0; \,\,\,\,\ i=1,...,q; \,\,\  j=q+1,...,d 
\end{equation}
Comparing these conditions with the statement of {\bf Proposition 2}, we conclude that the invariance of subspace ${\cal Q}$, under the action of the super-operator ${\bf {\bar U}}$, is a necessary condition for the existence of an operator  ${\bf {\bar U}}^{A} \in {\cal L}_{{\cal H}_2^{n_A}}$, obeying eq.(\ref{condition}). In such case ${\bf {\bar U}}^{A}={\bf {\bar U}^{QQ}}$, and the matrix $ {\bf {\bar U}^{QQ} }= [{\bar u}_{ij}^{QQ} ]_{i=1,...,q; j=1,...,q } $ is unitary. 

The following proposition summarizes these results.

{\bf Proposition 4.}   {\it Let ${\cal K} \subset {\cal L}_{{\cal H}_2^ n} $ to be the kernel of the linear partial trace transformation ${\bf {\bar T}} : {\cal L}_{{\cal H}_2^n} \rightarrow {\cal L}_{{\cal H}_2^{n_A}}$. ${\cal K}$ is invariant under the unitary operator ${\bf {\bar U}}:{\cal L}_{{\cal H}_2^n} \rightarrow {\cal L}_{{\cal H}_2^n}$ iff ${\bf {\bar U}^{KQ} }=0 $ ( or ${\bf {\bar U}^{QK} }=0)$. Then ${\bf {\bar U}}={\bf {\bar U}^{QQ} } \bigoplus {\bf {\bar U}^{KK} }$ and there is a unitary transformation ${\bf {\bar U}^{A} }= {\bf {\bar U}^{QQ} }:  {\cal L}_{{\cal H}_2^{n_A}} \rightarrow {\cal L}_{{\cal H}_2^{n_A}}$, such that, for every ${\bf Z} \in  {\cal L}_{{\cal H}_2^n} $, $ {\bf {\bar U}^{A}}{\bf {\bar T}}({\bf Z})={\bf {\bar T}}{\bf {\bar U}}({\bf Z}) $.}

\subsection{Condition for dimension reduction}

Applying the transformation ${\bf {\bar T}} =Tr_B \{. \}$ to each sate of the state set $ Q= \{ \rho_w = {\bf {\bar U}_w }(\rho_0)  : w \in \Sigma^{*}\}$ of the QA ${\cal M}$, we obtain the set $ Q_A= \{ \rho_w^{A}= Tr_B\{ \rho_w \} : w \in \Sigma^{*} \}$, where $\rho_w^{A}$ are the reduced density operators in ${\cal L}_{{\cal H}_{2}^{n_A}}$. In particular, the initial state $\rho_0$ is mapped into $\rho_0^{A}= Tr_B \{\rho_0 \}$. The set $Q_A$ will be a set of reachable states of a QA ${\cal M}_A$, iff there is a family of operators $ {\cal U}_A = \{ {\bf U_w^{A}}: w \in \Sigma^{*} \}$, such that $ \rho_w^A = {\bf  U_w^A} \rho_0^A {\bf  U_w^{A \dag}} $. We have shown in {\bf Proposition 4} that when ${\bf {\bar U}_{\sigma}^{KQ}}=0 $, then ${\bf {\bar U}_{\sigma}}={\bf {\bar U}_{\sigma}^{QQ}} \bigoplus {\bf {\bar U}_{\sigma}^{KK}}  $ and there is a unitary operator ${\bf {\bar U}_{\sigma}^{A} }={\bf {\bar U}_{\sigma}^{QQ}}$. By {\bf Proposition 3},  ${\bf {\bar U}_w}={\bf {\bar U}_w^{QQ}} \bigoplus {\bf {\bar U}_w^{KK}}  $ ($w \in \Sigma^{*})$ iff ${\bf {\bar U}_{\sigma}^{KQ}}=0 $, for every $\sigma \in \Sigma$.

Moreover, the quantum automata ${\cal M}$ and ${\cal M}_A$ will be equivalent, if they have the same behavior. The probability of obtaining an outcome $a_k$ when the QA $ { \cal M}$ is in the state $\rho_w$ is: $\omega (\rho_w, a_k) =Tr \{ {\bf {\tilde P}}_k(1) \rho_w \} = Tr_A\{ {\bf {\tilde P}}_k(1) \rho_w^{A} \}= \omega_A(\rho_w^{A}, a_k)$, where we used eq.(\ref{reduced}). This last equality shows that the quantum automata ${\cal M}$ and ${\cal M}_A$ have the same behavior: $\beta_{\cal M} (w) = \beta_{{\cal M}_A} (w): \Sigma^{*} \rightarrow \Omega $, for every $w \in \Sigma^{*}$.

The results that we have been deriving can be summarized in the following theorem.

{\bf Theorem 2:} {\it A tuple $ {\cal M}_A= \langle {\cal H}_2^{n_A}, \rho_0^{A}, \,\ \Sigma, \,\ {\bf {\bar {\cal U}}}_A, \,\ Q_A, {\tilde {\bf A}}(1),  \Omega \rangle   $, with  $ n_A < n$ and whose state set $ Q_A = \{  \rho_{  w }^{A}  = Tr_B \{ { \bf {\bar U }}_{ w }  (\rho_0)\}: w \in \Sigma* \} $ is a QA equivalent to the QA  ${\cal M}$, iff ${\bf {\bar U}_{\sigma}^{KQ}}=0$ for every unitary operator  ${\bf {\bar U}}_{\sigma \in \Sigma}$. In such a case, ${\bf {\bar U}_{\sigma}}={\bf {\bar U}_{\sigma}^{QQ}} \bigoplus {\bf {\bar U}_{\sigma}^{KK}}$ : \\
1 - $ {\cal H}_2^{n_A} $ is the underlying Hilbert space with dimension $2^{n_A} <2^ n$.\\
2 - $ \rho_{ 0 }^{A} =Tr_B \{ \rho_0 \} \in  {\cal L}_{{\cal H}_{2}^{n_A}}$ is the initial state.\\
3 - The input alphabet, $\Sigma = \{ \sigma \}$, is the same as in QA ${\cal M}$.\\
4-  The state set $ Q_A = \{ \rho_w^{A} : w \in \Sigma^{*} \} $ with $ \rho_w^{A} = Tr_B \{ \rho_w \} = {\bf {\bar U}_{\sigma}^{QQ}}( \rho_w)= {\bf { \bar U}_w^{A}} ( \rho_0^{A} )= {\bf  U}_w^{A} \rho_0^{A} {\bf  U}_w^{A \dag}$, where $\{ { \bf U}_w^{A} \}_{w \in \Sigma^*}$ is a family of unitary operators of ${\cal H}_2^{n_A}$ built as follows: i) ${\cal U}_A= \{ {\bf U_{\sigma}^{A}}  \}_{\sigma \in \Sigma}$, is a finite set of unitary operators indexed upon $ \sigma \in \Sigma  $, ii) ${ \bf U}_{\epsilon}^{A}={\bf 1}(A) $;  and iii) ${\bf U}_{ \sigma w}^{A}= {\bf  U}_{\sigma}^{A} {\bf U}_w^{A}$.\\
5 - The transition map $ \delta_A : \Sigma \times Q_A \rightarrow Q_A $ such that $ \rho_{w \sigma}^{A} = \delta_A (\sigma, \rho_w^{A} ) ={ \bf {\bar U }_{\sigma }^{A}} ( \rho_w^{A}) $. \\ 
6 - The output function $ \omega_A : Q \rightarrow \Omega $ is defined by $ \omega_A (\rho_w^{A},a_k) = Tr_A \{ {\bf { \tilde P}}_k (1) \rho_w^{A} \} = \omega(\rho_w,a_k) $.\\
7 - The output set, $ \Omega = \{ \Omega_w; w \in \Sigma^{*} ; \,\  \sum_k  \omega_A ( w, a_k) =1\}$, with $\,\ \Omega_w= \{\omega_A ( w, a_k):  a_k \in Spec {\bf A}\}$, is the same as in QA  ${\cal M}$. }

This theorem tells us how to built the physical support of the equivalent automaton ${\cal M}_A$. It is a system of $n_A ( n_1 < n_A < n)$ two-state quantum particles prepared in a quantum state $ \rho_{ 0 }^{A} =Tr_B \{ \rho_0 \}$, and submitted to quantum gates represented by the set of unitary operators ${\cal U}_A= \{ {\bf U}_{\sigma}^{A} \}_{\sigma \in \Sigma}$ such that ${\bf {\bar U}^{A}}_{\sigma}(\rho_w^{A}) = {\bf U}_{\sigma}^{A} \rho_w^{A} {\bf U}_{\sigma}^{A \dag}$ and ${\bf {\bar U}^{A}}_{\sigma}={\bf  {\bar U}_{\sigma}^{QQ}}$. The probabilities of the possible outcomes of a measurement of the observable ${\bf {\tilde A}}(1)= {\bf A}(1) \otimes {\bf 1}(B)$, performed on the first $n_1$ of the $n_A$  two-state particles, give the output symbols of the smaller QA ${\cal M}_A$, with dimension $2^{n_A} \times 2^{n_A}$. 

If there is at least one input $\sigma \in \Sigma$ such that ${\bf {\bar U}_{\sigma}^{KQ}} \neq 0$, then there is no QA with dimension $2^{n_A} \times 2^{n_A}$, equivalent to $ {\cal M}$.

\section{Minimization algorithm}

{\bf Theorem 2} says that a QA ${\cal M_A}$, working with $n_A <n$ is  equivalent to the initial QA ${\cal M}$ iff the kernel of the partial trace transformation is invariant under all the unitary operators ${\bf  {\bar U}_{\sigma}}$, with $\sigma \in \Sigma$. However, it does not say whether $n_A$ is the minimum number of qubits ${\bar n}$. In this section we present an algorithm to determine the minimal QA.

Usually, the density matrices $\rho$ and the unitary operators ${\bf U}_{\sigma}$, are expressed in terms of the computational Dirac basis-$ {\cal B}_d^C (n)$ of ${\cal L}_{{\cal H}_2^{n}}$, (Appendix A), rather than in terms of the basis-${\cal B}^{\cal Q} \cup {\cal B}^{\cal K}$, that is needed to check the invariance of subspace $\cal K$ under the operators ${\bf {\bar U}}_{\sigma}$. Therefore, the first step of the algorithm consists in constructing two ortonormal basis sets for the subspaces ${\cal Q}$ and ${\cal K}$, as explained in Appendix A. 

We assume that the $n_1$ first qubits of the QA ${\cal M}$ belong to subsystem (1) where the observable $ { \bf {\tilde  A}}(1)$ is going to be measured and that we have already checked if the conditions of {\bf Theorem 1} are verified. If so we obtain the corresponding sober  automaton.

 Given a QA $ {\cal M}= \langle {\cal H}_2^{n}, \rho_0, \Sigma,  {\cal U}, Q, {\bf {\tilde A}}(1), \Omega \rangle $ over a finite signature $\langle \Sigma, \Gamma \rangle $, we compute the QA ${\bar {\cal  M} }= \langle {\cal H}_2^{\bar n},{\bar \rho}_0, \Sigma,  {\bar {\cal U}}, {\bar Q}, {\bf {\tilde A}}(1), \Omega \rangle $, working with the minimal number of qubits ${\bar n}$, as follows:

1. Replace each Dirac vector $  \bigotimes_{j=1}^{n} v_{i_{j} i_{j}^{'}} ; i_{j}, i_{j}^{'} =0,1$, of the Dirac computational basis set ${\cal B}_d^{C}(n)$ by the $2^n \times 2^n$ matrix $ {\bf B_r} (r=2^n(l-1)+l'$), whose elements are all null except the element $b_{l,l^{'}}$, that is equal to $1$ and obtain the ordered computational basis set ${\cal B}_d^{C}(n)=\{ {\bf B_r}, r=1,...,2^n\times 2^n \}$.

2. Do $n_A=n_1$.

3. Replace each Dirac vector $  \bigotimes_{j=1}^{n_A} v_{i_{j} i_{j}^{'}} \in  {\cal B}_d^{C}(n_A); i_{j}, i_{j}^{'} =0,1$, of the Dirac computational basis set ${\cal B}_d^{C}(n_A)$, by the $2^{n_A} \times 2^{n_A}$ the matrix $ {\bf B_r^I},(r=2^{n_A} (l-1)+l')$, whose elements are all null except the element $b_{l,l^{'}}$, that is equal to $1$ and obtain the ordered computational basis set ${\cal B}_d^{C}(n_A)=\{ {\bf B_r^{' I}}, r=1,...,2^n_A \times 2^n_A \}$.

4. Compute ${\bf {\bar T}}({\bf B_r}) $ for $r=1,...,2^n \times 2^n$. If ${\bf {\bar T}}({\bf B_r})=0 $ add the vector ${\bf B_r}$ to the set $S_0$. If ${\bf {\bar T}}({\bf B_r})= {\bf B_j^{' I}}$, with $j=1,...,2^{n_A} \times 2^{n_A}$, add ${\bf B_r}$ to the set $S_j$.

5. With vectors of each set $S_j = \{ {\bf B_{r_k}}(j): {\bf {\bar T}} ( {\bf B_{r_k}}(j))= {\bf B_j^{' I}}; \,\ k=1,..., 2^{n-n_A} \}$ , $j=1,...,2^{n_A}\times2^{n_A}$, obtain the following linear combinations: ${\bf B_j^Q} = \frac{1}{\sqrt {2^{n-n_A}}}  \sum_{k=1}^{2^{n - n_A} } {\bf B_{r_k}}(j)$ and build the set ${\cal B}^{\cal Q} = \{ {\bf B_j^Q}: j=1,...,2^{n_A}\times 2^{n_A} \}$. 

6. With the vectors of each set $S_j$, $j=1,...,2^{n_{A}} \times2^{n_{A}}$, obtain the following linear combinations: $ {\bf B_{1j}} =  \sum_{k=1}^{2^{n - n_A} }c_k  {\bf B_{r_k}}(j)$ where $c_k =\frac{1}{\sqrt{2^{n-n_{A}}-1} } $ for $k=1,...2^{n-n_{A}}-1$ and $ c_{2^{n-n_{A}}}=-1$. 

7. Compute the vectors ${\bf B_{lj}}$ corresponding to all cyclic permutations of the coefficients $c_k$ ($k>1$) in eq.(\ref{LC2}), and build the $2^{n_{A}} \times 2^{n_{A}}$ sets $C_j= \{ {\bf B_{lj}}; l=1,...,2^{n-n_{A}}-1 \}$ with $j=1,...,2^{n_{A}} \times 2^{n_{A}}$.

8. Apply the Gram-Schmidt algorithm to each set $C_j$ in order to obtain a set  of orthonormal vectors $C_j^{\perp}$. 

9. Apply an appropriate ordering algorithm to the vectors given by (\ref{basisK}) to obtain an ordered set of orthonormal vectors ${\cal B^K} = \{ {\bf B_r^K}; r=2^{n_A} \times 2^{n_A}+1,...,2^n \times 2^n \}$ and  build ${\cal B}^C  ={\cal B}^{\cal Q} \cup {\cal B}^{\cal K }$. 

10. Write the transition matrix ${\bf {\bar C}}$ from the basis-${\cal B}_d^{C}(n)$ to the basis-${\cal B}^C$.

11. For each ${\bf U}_{\sigma}^{'} \in { \cal U}$ compute ${\bar u}_{ij}(\sigma) = Tr \{ {\bf B_j^{C }} {\bf U}_{\sigma}^{' }{\bf B_i^{C \dag}}{\bf U}_{\sigma}^{' \dagger} \}$ with $i,j=1,...,2^n \times 2^n$.

12. Build the matrices ${\bf {\bar U}}_{\sigma}^{'} =[ {\bar u}_{ij}(\sigma)] $, $i,j=1,...,2^n \times 2^n$, $\sigma \in \Sigma$.
 
13. For each $\sigma \in \Sigma$ compute ${\bf {\bar U}}_{\sigma} = {\bf {\bar C}}^{-1} {\bf {\bar U}}_{\sigma}^{'}{\bf {\bar C}}$

14. For each $\sigma \in \Sigma$ compute ${\bar u}_{ij}^{QK}(\sigma) = Tr \{ {\bf B_i^{K \dag}} {\bf U}_{\sigma} {\bf B_j^{Q }}  {\bf U}_{\sigma}^{\dagger} \}$, $i=2^{n_A} \times 2^{n_A}+1,...,2^n \times 2^n$ and $j =1,...,2^{n_A} \times 2^{n_A}$. 

15. If there is at least one
${\bar u}_{ij}(\sigma)^{QK} \neq  0$, go  to 17.  If for every $\sigma \in \Sigma$ and for every $i=1,...,2^{n_A} \times 2^{n_A},j=2^{n_A} \times 2^{n_A}+1,...,2^n \times 2^n$, ${\bar u}_{ij}^{QK}(\sigma) =  0$, go to 16.

16. For each $\sigma \in \Sigma$ build the matrices ${\bf {\bar U}_{\sigma}^{QQ}}= [{\bar u}_{ij}^{QQ}( \sigma )]=Tr \{ {\bf {\bar  U}}_{\sigma}^{\dagger}( {\bf B_j^{Q }}) {\bf B_i^{Q}} \}; i,j=1,...,2^{n_A} \times 2^{n_A}$, go to 20.

17. Do $n_A=n_A+1$.

18. If $n_A=n$, go to 21. If not, go to 3.

19. {\bf The initial QA ${\cal M}$, is the minimal one}.

20. Write ${\bar n}= n_A$.

21. Compute ${\bar {\rho_0}} = Tr_B\{ \rho_0\}$.

22. Write the set ${\bf {\bar {\cal U}}} = \{ {\bf {\bar U}_{\sigma}^{QQ}} ; \sigma \in \Sigma \}$.

23. {\bf The QA $ {\bar {\cal M}}= \langle {\cal H}_2^{\bar n}, {\bar {\rho_0}}, \Sigma, {\bf {\cal U}}, {\bar Q}, {\tilde {\bf A}}(1), \Omega \rangle   $, with  ${\bar n} < n$ qubits, is the minimal one}.

The complexity of this algorithm is computed in terms of the dimension of the initial QA ${\cal M}$, and of the dimension of the QA ${\cal M}_A$ which are $dim[ {\cal M}]= 2^n \times 2^n =N$ and $dim[ {\cal M}_A]= 2^{n_A} \times 2^{n_A} =N_A$ respectively. Another variable playing a role in the computation of the complexity of the algorithm, is the size $|\Sigma |$ of the input alphabet, since it gives the number of unitary operators ${\bf U}_{\sigma}$ (gates) to be computed in the ${\cal B}^C  ={\cal B}^{\cal Q} \cup {\cal B}^{\cal K }$-basis. The cardinality of the state set $ Q$, doesn't play any role in the minimization algorithm. The dimensions of the observable ${\bf {\tilde A}}(1)$ and the size $|\Omega|$ of the output set, are irrelevant variables in the algorithm.

{\bf Theorem 3:} {\it Given a QA $\cal M$ over a signature $( \Sigma, \Omega )$, the minimization algorithm requires $O(p(|\Sigma|, dim[{ \cal M}]))$ arithmetic operations.}

{\bf Proof:} Here is the detailed analysis of the algorithm (bounding for each step the worst case execution time in terms of arithmetic operations):

1 - $O(N^2)$ since the cost of  ordering $N$ objects is quadratic and the cost of computing the matrix representation of $N$ vectors $  \bigotimes_{j=1}^{n} v_{i_{j} i_{j}^{'}} ; i_{j}, i_{j}^{'} =0,1$ is $N$ .

2 - $O(1)$ since this is the cost of a constant.

3 -  $O(N_A^2)$ since the cost of  ordering $N_A$ objects is quadratic and the cost of computing the matrix representation of $N_A$ vectors $  \bigotimes_{j=1}^{n_A} v_{i_{j} i_{j}^{'}} ; i_{j}, i_{j}^{'} =0,1$ is $N_A$ .

4 - $O(N^{3/2}N_A^{-1/2})$ since the body of the cycle is running $O(N)$ times and each run computes  the partial trace ${\bf {\bar T}}\{ {\bf B_r} \}$ with the cost $O((N/N_A)^{1/2})$

5 - $O(N)$ since this is the cost of a sum of vectors of dimension $N$.

6 - $O(N)$ for the same reason as in 5.

7 - $O((NN_A)^{1/2})$ since $O((N/N_A)^{1/2})$ is the cost of a permutation of $(N/N_A)^{1/2}$ objects and there are $N_A$ permutations to compute.

8 - $O(N^{3/2} N_A^{1/2})$ since $O(N^2(N/N_A)^{1/2}$ is the cost of applying the Gram-Schmidt process to a set of $(N/N_A)^{1/2}$ vectors of dimension $N$ each, and this process is repeated for each of the $N_A$ sets $C_j$.

9 - $O((N-N_A)^2)$ since the cost of the ordering algorithm is quadratic in the number of objects and this is the number of vectors in the set given by eq.(\ref{LC1}).

10 - $O(N^2)$ since each of the $N$ vectors of the ${\cal B}_d^C$-basis has $N$ coefficients, being $N^2$ the dimension of the super-operator ${\bf { \bar C}}$.

11 - $O(N^5)$ since there are $N^2$ operations corresponding to two loops, being the cost of the body of each loop $O(N^3)$. The body of each loop computes the matrix product $ \{ {\bf B_j^{C }} {\bf U}_{\sigma}^{' }{\bf B_i^{C \dag}}{\bf U}_{\sigma}^{' \dagger} \}$  that has a cubic cost, followed by the computation of its trace that has cost $O(N)$. 

12 - $O(|\Sigma| N^5)$ since the body of the cycle is running $O(\Sigma|)$ times and the cost of each run is $O(N^5)$.

13 - $O(| \Sigma| N^6)$ since the body of the cycles is running $O(\Sigma|)$ times and the cost of each run is cubic (product of matrices) in the dimension of the super-operators ${\bf {\bar C}}$ which is $N^2$.

14 - $O(|\Sigma| (N-N_A)N^3)$ the same reason as in step 12.

15 - $O(|\Sigma| N^2)$ since the body of the cycles is running $O(|\Sigma| N^2)$ times and the cost of each run is $O(1)$.

16 - $O((|\Sigma| N_AN^3)$ the same reason as in step 12.

17 - $O( |\Sigma| log(N/N_A) N^6))$ since the body of the cycle is running $O( log(N/N_A) )$ times and the cost of each run is $O(| \Sigma| N^6)$.

18, 19 and 20 - $O(1)$ since this is the cost of a constant.

21 - $O(N^{3/2}N_A^{-1/2})$, the same reason as in step 4.

22 and  23 - $O(1)$ since this is the cost of a constant.

In conclusion: all steps of the minimization algorithm requires a polynomial number of arithmetic operations in $|\Sigma|$, $N= dim[{\cal M}]$, $N_A= dim[{\cal M}_A]$ and the $dim[{\cal M}_A] \leq dim[{\cal M}]$. {\bf QED}.

\section{Conclusion}

In this paper a new model for a QA working with qubits was proposed and the problem of minimizing its dimension was solved. 

The quantum states of the QA were represented by {\it density operators} which is a powerful approach to deal with measurements concerning only part of a quantum system.  The linearity of the automaton and of the partial trace super-operator were used to derive the conditions for dimension reduction.  It was shown that this is possible, if the kernel of the partial trace transformation is invariant under each of the unitary transformations associated to each letter of the input alphabet. 

It was also developed a minimization algorithm and it was shown that its complexity is polynomial in the size of the input alphabet and in the dimension of the QA.

Let us stress again that the minimization of the dimension o a QA and the minimization of the cardinality of its state set are different issues. While the dimension minimization can be addressed whether the QA is  finite or not, the cardinality minimization is only possible if the automaton is a QFA. This problem will be studied in a future work.

Finally, we refer the possibility that  the minimization technique here developed, can be adapted to other kind of automata, namely  Quantum Cellular Automata that implement quantum computation on qubits using spins \cite{WERNER2006, SCHUMACHER2004}.

\section*{Appendix A}

The computational basis set for ${\cal L}_{{\cal H}_2^{n}}$ is  an orthonormal basis set. In the Dirac notation it is expressed by ${\cal B}_d^{C}(n)= \{  \bigotimes_{k=1}^{n} v_{i_{k} i_{k}^{'}} ; i_{k}, i_{k}^{'} =0,1\}$ with $v_{i_{k} i_{k}^{'}} =  |i_k \rangle \langle  i_k^{'} | $  and $dim[ {\cal B}_d^{C}(n) ] =2^n \times 2^n$. 
To order the Dirac vectors of this basis we associate to each Dirac vector $  \bigotimes_{k=1}^{n} v_{i_{k} i_{k}^{'}} \in  {\cal B}_d^{C}(n); i_{k}, i_{k}^{'} =0,1$, the $2^n \times 2^n$ matrix $ {\bf B_r}$, whose elements are all null except the element $b_{l,l^{'}}$, that is equal to $1$ and where $r=2^n (l-1)+l'$.  Applying the same reasoning to the Dirac vectors $  \bigotimes_{k=1}^{n_A} v_{i_{k} i_{k}^{'}} \in  {\cal B}_d^{C}(n_A); i_{k}, i_{k}^{'} =0,1$, (which is the Dirac computational basis for the image space ${\cal L}_{{\cal H}_2^{n_A}}$) we associate to each vector $  \bigotimes_{j=1}^{n_A} v_{i_{j} i_{j}^{'}} \in  {\cal B}_d^{C}(n_A); i_{k}, i_{k}^{'} =0,1$,  a $2^{n_A} \times 2^{n_A}$ matrix $ {\bf B_r^I}$, whose elements are all null except the element $b_{l,l^{'}}$, that is equal to $1$ and where $r=2^{n_A} (l-1)+l'$.  Call ${\cal B}^I = \{ {\bf B_r^{' I}}, r=1,...,2^{n_A}\times 2^{n_A} \}$ to the ordered computational basis-set of the image space ${\cal L}_{{\cal H}_2^{n_A}}$.

The image of any vector of the basis-${\cal B}_d^{C}(n)$ by the transformation ${\bf {\bar T}} = Tr_B \{.\}$ is,
\begin{equation}\label{dirac}
{\bf {\bar T}} \left( \bigotimes_{k=1}^{n_A} v_{s_{k} s_{k}^{'}} \bigotimes_{k=n_A+1}^{n} v_{s_{k} s_{k}^{'}} \right)= \bigotimes_{k=1}^{n_A} v_{s_{k} s_{k}^{'}}Tr_B \{ \bigotimes_{k=n_A+1}^{n} v_{s_{k} s_{k}^{'}}\}
\end{equation}
Computing the partial trace we obtain,

\[ {\bf {\bar T}} \left( \bigotimes_{k=1}^{n_A} v_{s_{k} s_{k}^{'}} \bigotimes_{k=n_A+1}^{n} v_{s_{k} s_{k}^{'}} \right)= \left \{ \begin{array}{ll}
0                                                                        & \mbox{if  $ \exists_j : s_k \neq  s_k^{'}$ } \\
\bigotimes_{k=1}^{n_A} v_{s_{k} s_{k}^{'}}={\bf B_j^{' I}}  & \mbox{if  $ \forall_k : s_k= s_k^{'}$ }
\end{array}
\right.  \]

Let us denote by $S_0 = \{ {\bf B_{r_k}}  :  {\bf {\bar T}} ( {\bf B_{r_k}}) =0; \,\,\ k=1,...,2^n (2^n-2^{n_A}) \}$ the set of all vectors of ${\cal B}_d^{C}(n)$ transformed by ${\bf {\bar T}}$ into zero. Let us denote by $S_j = \{ {\bf B_{r_k}}(j): {\bf {\bar T}} ( {\bf B_{r_k}}(j))= {\bf B_j^{' I}}; \,\ k=1,..., 2^{n-n_A} \}$ the set of the vectors of $ {\cal B}_d^C (n)$ with the same image ${\bf B_j^{' I}}$. There are $2^{n-n_A}$ vectors ${\bf B_{r_k}}(j) \in {\cal B}_d^C (n)$ and there are $2^{n_A}\times 2^{n_A}$ sets $S_j $. 

A straightforward way of obtaining an orthonormal basis set for the subspace ${\cal Q}$ consists in taking the following linear combinations of all vectors ${\bf B_{r_k}}(j) \in S_j$:
\begin{equation}\label{LC1}
{\bf B_j^Q} = \frac{1}{\sqrt {2^{n-n_A}}}  \sum_{k=1}^{2^{n - n_A} } {\bf B_{r_k}}(j)
\end{equation}
The set ${\cal B}^{\cal Q} = \{ {\bf B_j^Q}: j=1,...,2^{n_A}\times 2^{n_A} \}$ is an ordered set of orthonormal vectors and constitute a basis set for the subspace ${\cal Q}$.  In fact, applying the super-operator ${\bf {\bar T}}$ to both sides of eq.(\ref{LC1}) the result is the vector ${\bf B_j^{I}}= \sqrt {2^{n-n_A}}{\bf B_j^{' I}}$.

With the vectors of the sets $S_j$ and $S_0$ it is possible to build up an orthonormal basis set for the kernel subspace ${\cal K}$, complementary to ${\cal Q}$. It is simple algebra to prove that the following linear combination of all vectors ${\bf B_{r_k}}(j) \in S_j$
\begin{equation}\label{LC2}
 {\bf B_{1j}} =  \sum_{k=1}^{2^{n - n_A} }c_k  {\bf B_{r_k}}(j)
\end{equation}
with $c_k =\frac{1}{2^{n-n_{A}}-1}  $ ; $k=1,...2^{n-n_{A}}-1$ and $ c_{2^{n-n_{A}}}=-1$, is applied on $0$ by the super-operator ${\bf {\bar T}}$.

Making all the cyclic permutations of the last $2^{n-n_A} -1 $ coefficients $c_k$ of eq.(\ref{LC2}) it is possible to generated $2^{n-nA} -1$ different vectors ${\bf B}_{lj} $ with $l=1,...,2^{n-nA} -1$. The vectors of each set $C_j= \{ {\bf B}_{lj} ; l=1,...,2^{n-nA} -1 \}$; $(j=1,...,2^{n_a}\times 2^{n_A})$, are orthogonal to the vectors of the basis-${\cal B^Q}$ and to the vectors of the set $S_0$. 

The vectors of each set $C_j $ are not mutually orthogonal. They  can be transformed in an set of orthormal vectors, $ C_j^{\perp}$, by applying to each set $C_j$ the Gram-Schmidt algorithm.
 
Giving all these properties of the vector sets $S_0$ and $C_j^{\perp}$, it is now clear that the following union of sets
\begin{equation}\label{basisK}
S_0 \cup_{j=1}^{2^{n_{A}} \times 2^{n_{A}}} C_j^{\perp}. 
\end{equation}
 is an ortonormal basis set for the kernel subspace ${\cal K}$. Applying an appropriate ordering algorithm to the vectors given by (\ref{basisK}) we obtain an ordered set of orthonormal vectors ${\cal B^K} = \{ {\bf B_r^K}; r=2^{n_A} \times 2^{n_A}+1,...,2^n \times 2^n \}$.

Let $ {\cal B}^C = \{ {\bf B_r}; r=1,...,2^n \times 2^n\}$ be the ordered orthonormal computational basis set in ${\cal L}_{{\cal H}_2^{n}}$.
This basis set is related to the orthonormal basis set $ {\cal B}^{\cal Q} \cup {\cal B}^{\cal K}$ through the equations 
\begin{equation}\label{changeQ}
{\bf B_i^Q}= \sum_{s=1}^{2^n \times 2^n} {\bar c}_{si}^Q {\bf B_s} ; \,\,\,\,\
{\bf B_i^K}= \sum_{s=1}^{2^n \times 2^n} {\bar c}_{si}^K {\bf B_s} 
\end{equation}
where ${\bar c}_{si}^Q ;\,\ ( i=1,...,2^{n_A} \times 2^{n_A} ; s= 1,...,2^n \times 2^n )$ and ${\bar c}_{si}^K \,\ ( i=2^{n_A} \times 2^{n_A}+1,..., 2^n \times 2^n;  s=1,..., 2^n \times 2^n)$ are the elements of the transition matrix ${\bf {\bar C}}$ from the basis-$ {\cal B}^C$ to the basis-$ {\cal B}^{\cal Q} \cup {\cal B}^{\cal K}$. The action of $ {\bf {\bar U}}$ relative to the $ {\cal B}^C$-basis is
\begin{equation}\label{changebyu}
{\bf {\bar U}}^{'}({\bf B_j})= \sum_{i=1}^{2^n \times 2^n}  {\bar u}_{ij}^{'}{\bf B_i}
\end{equation}
where ${\bar u}_{ij}^{'}=({\bf {\bar U}}^{'}({\bf B_j}),{\bf B_i})= Tr\{ {\bf {\bar U}}^{'  \dagger} ({\bf B_j}){\bf B_i} \} $ are the elements of ${\bf {\bar U}}$ relative to this basis set.
If ${\bf {\bar U}}^{'}$ is the matrix of the super-operator ${\bf {\bar U}}$ relative to the  $ {\cal B}^C$-basis, then  
\begin{equation}\label{changeU}
{\bf {\bar U}} = {\bf {\bar C}}^{-1} {\bf {\bar U}}^{'}{\bf {\bar C}}
\end{equation}
is the matrix of the same super-operator relative to the basis-$ {\cal B}^{\cal Q} \cup {\cal B}^{\cal K}$. The matrix ${\bf {\bar U}}$ given by eq.(\ref{changeU}), is the one that must be used to compute the matrices ${\bf {\bar U}^{QQ}}$ and ${\bf {\bar U}^{KQ}}$ of eqs.(\ref{block11}) and (\ref{block21}).


\newpage

\begin{thebibliography}{99}

\bibitem{MOORE}

C. Moore and J. Crutchfield. Quantum automata and quantum grammars. {\it Theoretical Computer Science}, {\bf 237}, pages 275-306 (2000).

\bibitem{WATROUS}

A. Kondacs and J. Watrous. On the power of quantum finite state automata. {\it In Proceedings of the 38th Annual Symposium on Foundations of Computer Science}, pages 66-75 (1997).

\bibitem{AMBAINIS98}

A. Ambainis and R. Freivalds, 1-way quantum finite automata: strenghts , weaknesses and generalizations. {\it In Proceedings of 39th Annual Symposium of Foundations of Computer Science}, pages 332-3341, (1998).

\bibitem{NISHIMURA}

H. Nishimura and T. Yamakami. An application of quantum finite automata to quantum proof systems. quant-ph/0410040.

\bibitem{AMBAINIS02}

A. Ambainis and A. Kikusts, Exact results for accepting probabilities of quanta automata, quant-ph/0109136v2.

\bibitem{BOOTH} T. L. Booth. {\it Sequential machines and automata theory}, John Wiley and Sons (1967).

\bibitem{WATROUS02}

A. Ambainis and J. Watrous. Two-way finite automata with quantum and classical states. {\it Theoretical Computer Science}, {\bf 287} (1), pages 299-311 (2002).

\bibitem{PAZ}

A. Paz. {\it Introduction to probabilistic automata}, Academic Press, New York (1971).

\bibitem{RABIN}

M. O. Rabin. Probabilistic Automata. {\it Info. Control}, {\bf 6}, pages 230-245 (1963).

\bibitem{KITAEV}

D. Aharonov, A. Kitaev and N. Nisan. Quantum circuits with mixed states,  {\it In Proceedings of the 30th Annual ACM Symposium on Theory of Computing}, pages 20-30 (1998), also in quant-ph9806029.

\bibitem{AMBAINIS}

A. Ambainis, A. Nayak, A. TA-Shma and U. Vazirani. Dense quantum coding and quantum finite automata, {\it In Proceedings of the 31th Annual ACM Symposium on Theory of Computing}, pages 376-383 (1999). Also in quant-ph/9804043 v2.

\bibitem{CAVES2000}

H. Barnum, C. M. Caves, C. A. Fuchs, R. Jozsa, and B. Schumacker. On quantum coding for ensembles of mixed states, quant-ph/0008024 (2000).

\bibitem{COHEN92}

C. Cohen-Tannoudji, B. Diu, F. Lalo\"{e}  {\sl Quantum Mechanics I},
Academic Press, New York (1996).

\bibitem{PRESKILL}

John Preskill, {\it Lecture Notes for Physics 229: Quantum Information and Computation}, California Institute of Technology. Available at http://theory.caltech.edu/people/preskill/ph229, (1998)

\bibitem{KRAUS}

K. Kellig and K. Kraus, {\sl Comm. Mathematical Phy. }, {\bf 16}, 142, (1970). M. A. Nielsen and I. L. Chuang {\sl Quantum Computation and Quantum Information}, Cambridge University Press, U.K. (2002).

\bibitem{HALMOS}

P. R. Halmos, {\sl Finite-dimensional vector-spaces}, Springer-Verlag, New-York (1987).

\bibitem{WERNER2006}

D. J. Shepherd, T. Franz and R. F. Werner, Phys. Rev. Lett. {\bf 97}, 020502 (2006).

\bibitem{SCHUMACHER2004}

B. Schumacher and R. F. Werner, quant-ph/0505174v1 (2004).


\end{thebibliography}
\end{document}